# Node-Based Job Scheduling for Large Scale Simulations of Short Running Jobs

Chansup Byun[1], William Arcand[1], David Bestor[1], Bill Bergeron[1], Vijay Gadepally[1,2],
Michael Houle[1], Matthew Hubbell[1], Michael Jones[1], Anna Klein[1], Peter Michaleas[1], Lauren Milechin[4],
Julie Mullen[1], Andrew Prout[1], Albert Reuther[1], Antonio Rosa[1], Siddharth Samsi[1], Charles Yee[1], Jeremy Kepner[1,2,3]
[1]MIT LLSC, [2]MIT CSAIL, [3]MIT Math, [4]MIT EAPS

*Abstract*—Diverse workloads such as interactive supercomputing, big data analysis, and large-scale AI algorithm development, requires a high-performance scheduler. This paper presents a novel node-based scheduling approach for large scale simulations of short running jobs on MIT SuperCloud systems, that allows the resources to be fully utilized for both long running batch jobs while simultaneously providing fast launch and release of large-scale short running jobs. The node-based scheduling approach has demonstrated up to 100 times faster scheduler performance that other state-of-the-art systems.

I. INTRODUCTION

Many supercomputing centers are witnessing a growth in the diversity of scientific workflows. Many centers are facing challenges to supporting both traditional supercomputing needs in addition to emerging technologies such as big data analysis and AI algorithm developments. For over a decade, the MIT Lincoln Laboratory Supercomputing Center (LLSC) has focused on developing unique, interactive, on-demand high-performance computing (HPC) environment to support diverse science and engineering applications. This system architecture has evolved into the MIT SuperCloud. The MIT SuperCloud has spurred the development of a number of cross-ecosystem innovations in high performance databases [1, 2], database management [3], data protection [4], database federation [5, 6], data analytics [7], dynamic virtual machines [8, 9], and system monitoring [10]. This capability has grown in many dimensions. MIT SuperCloud not only continues to support parallel MATLAB and Octave jobs, but also jobs in Python [11], Julia [12], R [13], TensorFlow [14], PyTorch [15], and Caffe [16] along with parallel C, C++, Fortran, and Java applications with various flavors of message passing interface (MPI) [17]. As the system capacity of the MIT SuperCloud has increased significant challenges have emerged enabling fast launches for interactive jobs while simultaneously supporting large background simulations.

Recent detailed studies [18, 19] comparing supercomputing schedulers and Big Data schedulers observed:

1. Supercomputing schedulers, including Slurm [20], GridEngine [21], and LSF [22], are capable of launching synchronously parallel (MPI-style) jobs as well as loosely parallel job arrays. Big Data schedulers including Mesos, Apache YARN [23], and the open-source Kubernetes project [24] only supported loosely parallel job arrays.

2. Slurm, Mesos, and Kubernetes were designed to handle 100,000+ jobs, both in its queues and executing on compute nodes.

Supporting complex and mixed scientific workloads at scale has spurred research into resource management and scheduling algorithms [18, 19, 25-27]. The MIT SuperCloud has developed a multi-level scheduling approach [18, 19, 28] to reduce scheduler overheads by aggregating multiple scheduling tasks into a single scheduling launch based on the available number of compute cores. This approach has been extended to aggregate all of the compute tasks assigned to a single node as a single scheduling task using a node-based scheduling approach [29]. Node-based scheduling, also termed "triples mode", has been integrated into MIT SuperCloud Matlab/Octave tools (pMatlab [30] and gridMatlab [31]) and general-purpose launch tools (LLsub and LLMapReduce [28]). Node-based scheduling has demonstrated launch of large scale interactive jobs at rate of over 5000 jobs/second (260,000+ Matlab/Octave processes in under 40 seconds) [29].

Fast launch requires available resources, but automatic preemption can be slow to terminate low-priority spot jobs to create these resources [32]. This results in a significant delay when launching an interactive job when preemption of spot jobs is required. The node-based scheduling approach can also be applied to preemptable spot jobs, allocating the compute resources for a given spot job by nodes instead of compute cores. Node based scheduling enables faster release of spot jobs and reduces the workloads on the scheduler, not only with spot jobs, but also with any other parallel or array jobs. Furthermore, node-based schedule of spot jobs allows for better usage of memory by holistically pinning processes to cores.

The inefficiencies of short running jobs are due to the overhead associated with the life cycles of the jobs such as scheduling, dispatching and cleaning up afterwards. Recently, there have been some efforts to overcome these challenges with various approaches such as packing many small jobs into a single large job using a special library called CRAM [33] and using multi-level scheduling [18, 19, 28] and hierarchical scheduling [25] approaches.

This material is based upon work supported by the Assistant Secretary of Defense for Research and Engineering under Air Force Contract No. FA8721-05-C-0002 and/or FA8702-15-D-0001. Any opinions, findings, conclusions or recommendations expressed in this material are those of the author(s) and do not necessarily reflect the views of the Assistant Secretary of Defense for Research and Engineering.

In this paper, we present the node-based scheduling approach to overcome the shortcomings of the traditional supercomputing schedulers with large numbers of short running jobs. In this approach, we have aggregated all of the compute tasks to be scheduled on the same node as a single scheduling task for the scheduler. This is accomplished using MIT SuperCloud-developed tools including LLsub, LLMapReduce, and pMatlab/gridMatlab. We compare the scheduler performance between the multi-level scheduling approach (LLMapReduce MIMO – Multi-Input, Multi-Output) and the node-based scheduling approach (LLMapReduce MIMO with the triples mode). We have demonstrated that the node-based scheduling approach has improved the scheduler performance significantly – up to 100 times increased performance compared to that of the multi-level scheduling approach.

## II. Approach

The multi-level scheduling approach, implemented in LLMapReduce MIMO, aggregates all the compute tasks to be executed on the same physical core as a single scheduling task by packing all individual compute tasks in a loop. This reduces the scheduling overhead due to repeated dispatching, loading and execution of multiple compute tasks. By doing so, it will launch an application only once and carry out the computation for all the tasks in a loop. With this approach, we have observed that it can increase system utilization over 90% for various types of scheduling software we have tested [1]. However, this approach has shown a performance limitation in the case of large numbers of short duration tasks. These tasks are become more prevalent as supercomputing systems are being deployed with an increasing number of compute nodes, and each compute node is becoming denser in terms of the number of cores per node.

We have applied a node-based scheduling approach in order to handle this scaling issue. By using a node-based scheduling approach, we can further reduce the number of scheduling events, and each scheduling tasks will have much longer runtimes if more than one short running tasks are aggregated and executed on each core. This, in turn, reduces the workload for the scheduler significantly.

This node-based scheduling approach generates a job execution script per each node on the fly in such a way that all of the compute tasks to be executed on the same node are aggregated as a single scheduling task for the scheduler. Because this aggregation is done explicitly and algorithmically, we can design how we want to manage the compute tasks. When generating the job submission script, we have also implemented explicit control of the process affinity and the number of threads of all the compute tasks. This provides an additional performance benefit with this node-based scheduling approach. This node-based scheduling approach was initially implemented in pMatlab/gridMatlab and then extended to other MIT SuperCloud tools (LLsub and LLMapReduce) using Slurm scheduler. However, the node-based scheduling approach is scheduler-agnostic and it can be used with most supercomputing schedulers.

## III. Performance Comparison

In the recent study performed on the MIT SuperCloud system [18], in order to compare the launch latency for the schedulers, we chose four representative schedulers from across the scheduler landscape: Slurm, Son of Grid Engine, Mesos, and Hadoop YARN. We evaluated the performance of the scheduler from the scheduler perspective. That is, we are executing constant time tasks to occupy a given job slot for a set amount of time. So, for that job slot, the scheduler does not need to reevaluate its assignment until that occupation time has expired. The constant time tasks are inserting delays after which the scheduler must schedule and dispatch a new task to a given resource.

TABLE I. PARAMETER SETS AND RUNTIMES USED TO MEASURE SCHEDULER LATENCY AS A FUNCTION OF JOB TASK TIME.

| Configuration | Rapid Tasks | Fast Tasks | Medium Tasks | Long Tasks |
|---|---|---|---|---|
| Task time, t | 1s | 5s | 30s | 60s |
| Job time per processor, $T_{job}$ | 240s | 240s | 240s | 240s |
| Tasks per processor, n | 240 | 48 | 8 | 4 |

Table 1 shows parameter sets and runtimes used to measure scheduler latency as a function of job task time in the previous study. We used the same parameter sets for the current study to measure the overall overhead for the life cycle of a job and compare the scheduler performance between the multi-level and node-based scheduling approaches. The short running jobs are categorized into four configurations: rapid (1 seconds), fast (5 seconds), medium (30 seconds) and long (60 seconds), respectively. Since we keep the job time per each processor constant, based on the configuration, the number of tasks per each processor varies from 240 to 4, as shown in Table I.

TABLE II. BENCHMARK CONFIGURATION

| Nodes | 32 | 64 | 128 | 256 | 512 |
|---|---|---|---|---|---|
| Cores per node | 64 | 64 | 64 | 64 | 64 |
| Processors, P (cores) | 2048 | 4096 | 8192 | 16384 | 32768 |
| Total processor time | 136.5 h | 273.1 h | 546.1 h | 1092.3 h | 2184.5 h |

In the current study, we conducted the performance comparison with 5 different scaling sizes, 32, 64, 128, 256 and 512 nodes, on the MIT SuperCloud system as shown in Table II. Each compute node has 64 physical cores. Since we create array jobs in such a way that each processor (core) performs a constant job time (240 seconds) the total processor times vary as shown in Table III, based on the configuration (number of nodes) sizes. The scale of problems considered in this study is much bigger than the one used in the previous study. The largest number of compute tasks from a job is almost 8 million when the rapid task (1 second) is deployed to fill up 512 nodes.

We compared the scheduler performance of the node-based scheduling approach with the current state-of-the-art multi-level scheduling approach using LLMapReduce. The node-based scheduling approach is an expansion of aggregation by node on top of the core-based aggregation done by the multi-level scheduling implementation in LLMapReduce MIMO. For example, the multi-level scheduling LLMapReduce MIMO creates an array job of the number of scheduling tasks equivalent to the number of processors, P, as shown in Table II whereas the node-level scheduling creates an array job of the number of scheduling tasks equivalent to the number of compute nodes by aggregating all the compute tasks to be performed on the same compute node as a single scheduling task.

*A. Cluster Systems*

The system we used for the study is the production system, TX-Green. The TX-Green system is made with many different types of compute nodes and now has nearly 70,000 cores available for users' parallel jobs. The majority of nodes are 648 Intel Xeon Phi compute nodes. Each node has a 64-core Intel Xeon Phi 7210 processor, for a total of 41,472 cores, along with 192 GB RAM, 16 GB of on-package MCDRAM configured in 'flat' mode, local storage, 10-GigE network interface, and an OmniPath network interface. Each compute node also has two local storage drives: a 128 GB solid state drive (SSD) and a 4.0 TB hard drive. The Lustre [34] central storage system is made up of two separate storage arrays: a 10 petabyte Seagate/Cray ClusterStor CS9000 storage array for sharing data in groups and a 14 petabyte DDN 14000 storage array for users' home directory, which are both directly connected to the core switch. Recently, we added an additional 225 compute nodes with 9,000 additional cores. Each of these nodes has two 20-core Xeon Gold 6248 [35] processors with 384 GB RAM, two Nvidia Volta V100 [36] GPUs with 32 GB RAM each, and a 25-Gigabit Ethernet network interface.

*B. Performance Measurements*

We measured the scheduler performance in two areas, scheduler overhead and system utilization. For each combination of benchmark configuration and task type as shown in Tables I and II, we measured the runtime three times for both scheduling approaches. The results are shown in Table III.

The job run time is defined as the time between the start time of the first task and the end time of the last task, shown in seconds. The M* and N* in Table III denote the multi-level and node-level scheduling approaches, respectively. The performance measurement of most of the jobs were conducted on a reserved resource for a given number of compute nodes matching the benchmark configuration on a production environment. However, for the multi-level scheduling jobs with 256 and 512 node configurations, we have to conduct all the benchmark runs on a dedicated environment right after a scheduled maintenance of the system. This is because the scheduler becomes very busy under heavy loads during the job submission and is unresponsive while clearing the finished tasks. This is not acceptable for a production environment. We only ran one task type (Long tasks) for 512 nodes configuration with the multi-level scheduling approach because it takes too long to release the completed tasks. More details are discussed in the next section.

In addition, there was an issue with the node state when running one of the runs for the medium tasks with the 256 node configuration. We found that it caused the job to be stuck in a pending state. We had to manually correct the issue in order to get the job started properly. Therefore, it resulted in a significantly longer runtime. However, we did not attempt to rerun the case because we only used the median values for studying the scheduling behavior.

Table III. Summary of Run Times. M* and N* denote the multi-level and node-level scheduling approaches.

| Task time, t | | 1 | 5 | 30 | 60 |
|---|---|---|---|---|---|
| 32 nodes | M* | 305, 284, 291 | 280, 278, 277 | 283, 284, 287 | 296, 283, 282 |
| | N* | 241, 242, 243 | 243, 242, 242 | 242, 242, 242 | 241, 242, 242 |
| 64 nodes | M* | 272, 291, 305 | 294, 293, 310 | 322, 298, 317 | 324, 317, 302 |
| | N* | 243, 242, 242 | 242, 242, 243 | 242, 242, 241 | 241, 242, 242 |
| 128 nodes | M* | 445, 424, 424 | 421, 427, 439 | 428, 424, 423 | 509, 435, 443 |
| | N* | 244, 255, 245 | 248, 245, 261 | 245, 251, 246 | 244, 254, 250 |
| 256 nodes | M* | 430, 429, 495 | 455, 453, 427 | 467, 474, 2464* | 442, 431, 452 |
| | N* | 256, 248, 257 | 272, 248, 247 | 252, 248, 247 | 244, 272, 251 |
| 512 nodes | M* | N/A | N/A | N/A | 2644, 2768, 2791 |
| | N* | 262, 391, 489 | 257, 254, 272 | 272, 264, 487 | 266, 487, 312 |

Fig. 1 shows the normalized overhead time for all the runs. This includes the combination of the four different types of tasks and 5 different benchmark configuration sizes. The plot was created by using the median value of the three runs from each combination of task types and benchmark configuration. The overhead time is calculated by subtracting the job time per processor, $T_{job}$, from the median runtime as shown in Table III. The horizontal axis represents the task times, which are made of four different times, 1, 5, 30, and 60 seconds, respectively. The vertical axis shows the overhead time, which is normalized with the job time per processor, $T_{job}$. The lower value means that there is less scheduler overhead in handling the life cycle of a job. Each benchmark configuration size is differentiated with different types of symbols as noted in Fig. 1, while the open and filled symbols denotes the multi-level and node-level scheduling approaches, respectively.

As shown in Fig. 1, it is clear that the node-based scheduling approach performs much better (less overhead time) than the multi-level scheduling approach for all the runs. The scheduler overhead with the node-based scheduling approach has been kept below 10% of the job time per processor, $T_{job}$, for most of the cases. There are four cases that exceed 10% of the job time but this may be influenced by the other jobs being served at the time since the runs are conducted under the production environment. The scheduler overhead with the multi-level scheduling approach exceeds 10% or more for all the runs.

In general, increasing the scale of a job for the given task time has also increased the scheduler overhead time for most cases. However, it is noted that the overhead time remains at the same level regardless of the task times for both cases, as long as the configuration size is kept the same. This shows that the scheduler overhead is dominated by the number of scheduling tasks since, in both cases, we are packing all tasks as per-core or per-node base for multiple level scheduling (LLMapReduce MIMO) and node-based scheduling (LLMapReduce MIMO with triples), respectively. With the 512 node configuration, the multi-node scheduling approach shows about 57x (based on the median runtime) and 100x (based on the best runtime) less overhead time as compared to the multi-node scheduling approach.

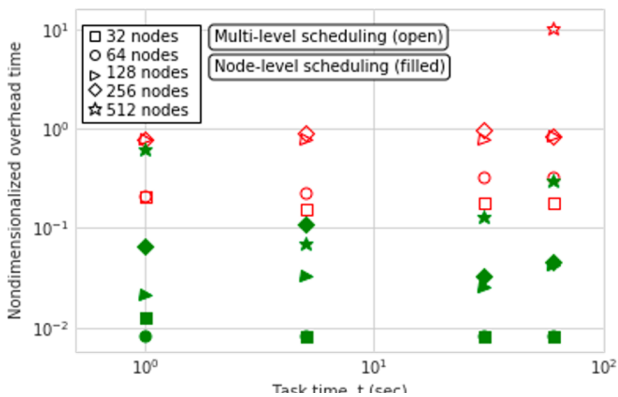

Fig. 1. Overhead time comparison for all the runs

Next, we looked at more details about how the scheduler behaves by looking at the scheduling events (start and end times) of all the scheduling tasks of each job from the scheduling log. We looked at only the cases that corresponds with the median values from all the runs as shown in Table III. Also, in order to compare the scheduling behavior between the multi-level and node-based scheduling approaches, we shifted the time in such a way that the initial time zero is to be the first scheduling event (start time) of the first task of each job from the scheduler log. This enables us to visualize how many tasks are dispatched to run and, in turn, how many resources are being used. It is noted that this does not include the latency between the time of job submission and the time the scheduler recognizes the submitted job.

We summarize all the plots in Fig. 2. This shows how much the system is being utilized by both the multi-level and node-level scheduling approaches for the given node configuration. If there is no overhead, each job will be dispatched and started instantly and the runtime of each run will be equal to the job time per processor, $T_{job}$. However, as shown in Fig. 2, we can clearly see that the multi-level scheduling approach exhibits significant delays to fully dispatch all the compute tasks to fill up the resources and fully utilize the compute resources 100%. Furthermore, for the 512 node configuration, it was unable to reach 100% system utilization at any point in time throughout its job life cycle. The node-based scheduling approach shows very little delay in starting the job and almost instantly achieves 100% utilization for almost all the runs.

As we scale up the job sizes, it takes longer to fully dispatch and utilize all the compute tasks with the multi-level scheduling approach. Interestingly, the cleanup of the completed tasks took even longer as the job sizes were scaled up. This increased the wall-clock time of those jobs, which impacts the time that compute resources are held, significantly. Especially with the 512-node configuration, it is noted that the scheduler becomes very busy most of the time while the job is being dispatched. Furthermore, it becomes unresponsive while cleaning up the completed tasks. It could not even dispatch some of compute tasks until a later stage (after the 2500 second mark) of the job. This is unacceptable behavior for the production systems because it will freeze all the jobs in the scheduling system and reduce the overall system utilization. This clearly shows the scaling issue with the multi-level scheduling approach.

However, the node based scheduling approach shows very little delay in dispatching compute tasks. All the compute tasks from each job have been scheduled and dispatched concurrently and achieved 100% system utilization instantly after the job submission. It also exhibits little overhead with cleaning up the completed tasks. Based on the plots, all the nodes are released approximately at the same time. It should be noted that all the runs with the node-based scheduling approach have been done under the production environment without affecting other users' jobs. But, for large configurations with 256- and 512-node configuration, a dedicated environment is needed in order to execute the multi-level scheduling runs, because the scheduler becomes unresponsive to handle the jobs. This clearly demonstrates the node-based scheduling approach is capable of handling very large simulation of short running jobs.

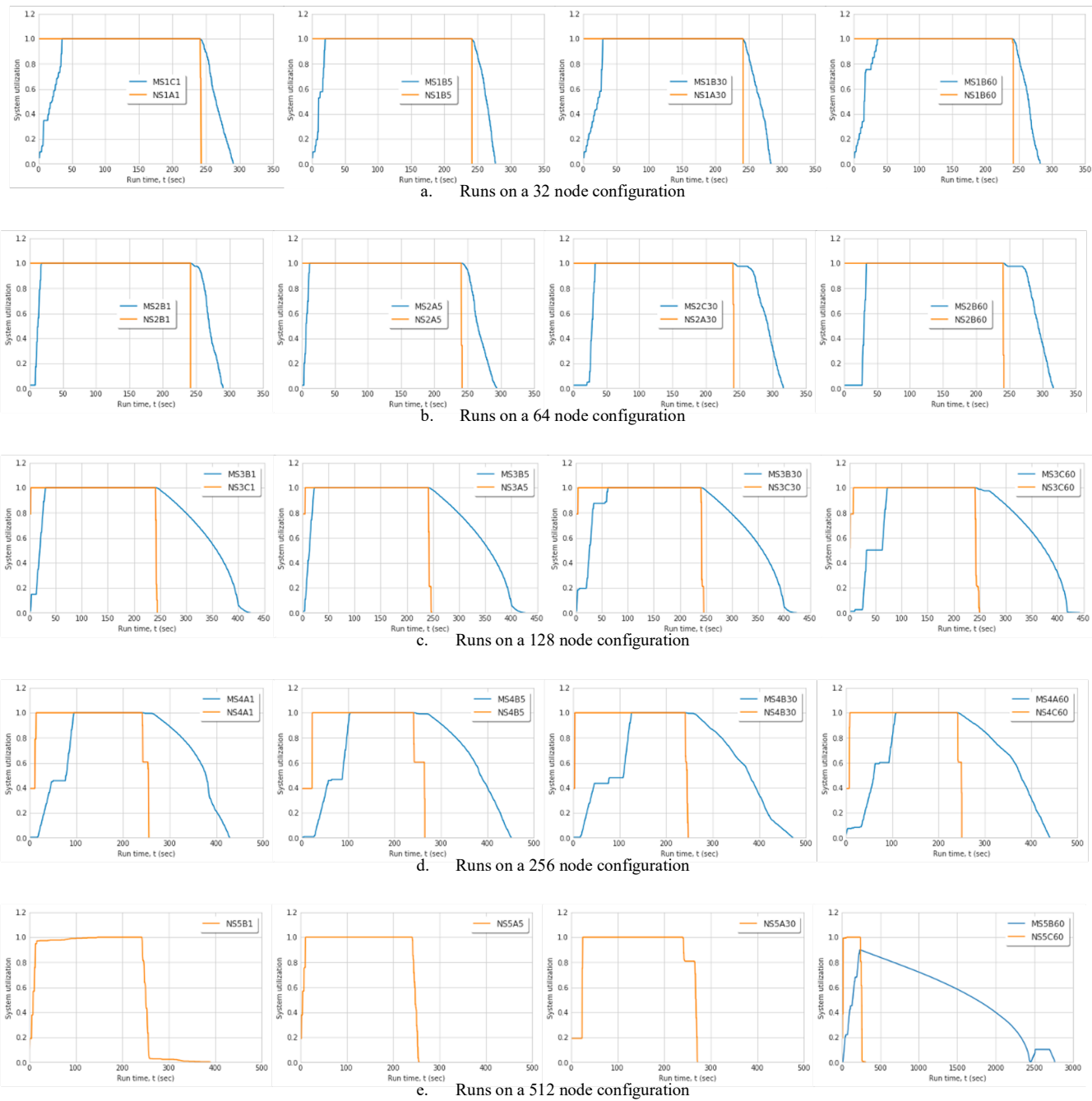

Fig. 2. System utilization over time for all the runs that corresponds with the median values in the Table III. For the legend label, M for multi-level scheduling, N for node-level scheduling, S1 (32 nodes) through S5 (512 nodes) configuration, A,B,C identifies three runs from each of the four task times (1, 5, 30, and 60 seconds)

## IV. SUMMARY

In this paper, we have demonstrated that the node-based scheduling approach (also called the triples mode with LLMapReduce MIMO) provides excellent performance in managing various types of jobs (relatively short running) regardless of its deployment sizes. We have also compared its scheduler performance with the multi-level scheduling approach (LLMapReduce MIMO) and demonstrated that the node-based scheduling approach provides much less overhead as compared to the multi-level scheduling approach. In addition, we have observed that the multi-level scheduling approach is showing decent performance at small size cases but shows significant performance degradation with 256 and 512 node size deployments, which have to rely on a dedicated system environment.

It is clear that, with the multi-level scheduling approach, as scaling up the problem sizes, the scheduler could not dispatch the compute tasks fast enough to fill up the compute resources. This results in poor system utilization over time and longer run times to complete the entire compute tasks. It is also interesting to note that releasing the completed tasks takes significantly longer as compared to dispatching the compute tasks to fill up the compute resources.

ACKNOWLEDGMENTS

The authors wish to acknowledge the following individuals for their contributions and support: Bob Bond, Alan Edelman, Jeff Gottschalk, Charles Leiserson, Dave Martinez, Steve Rejto, Marc Zissman and Lincoln Laboratory Supercomputing Center postdocs, Nathan Frey, Hayden Jananthan, Joseph McDonald, Matthew Weiss.